\documentstyle[aps,manuscript]{revtex}  
\begin{document} 
\title{Directed Paths on Hierarchical Lattices with Random Sign
Weights} 
\author{Eduardo G. Aponte} 
\address{Laboratorio de F\'\i sica
Computacional UNEXPO, Torre Domus 14-B, Caracas, Venezuela}
\author{Ernesto Medina} 
\address{Laboratorio de F\'\i sica Estad\'\i
stica de Sistemas Desordenados, Centro de F\'\i sica, IVIC,\\ Apartado
21827, Caracas 1020A, Venezuela.}  

\date{\today}
\maketitle

\begin{abstract}
We study sums of directed paths on a hierarchical lattice where each
bond has either a positive or negative sign with a probability
$p$. Such path sums $J$ have been used to model interference effects
by hopping electrons in the strongly localized regime. The advantage
of hierarchical lattices is that they include path crossings, ignored
by mean field approaches, while still permitting analytical
treatment. Here, we perform a scaling analysis of the controversial
``sign transition'' using Monte Carlo sampling, and conclude that the
transition exists and is second order. Furthermore, we make use of
exact moment recursion relations to find that the moments $\langle
J^n\rangle$ always determine, uniquely, the probability distribution
$P(J)$. We also derive, exactly, the moment behavior as a function of
$p$ in the thermodynamic limit. Extrapolations ($n\rightarrow 0$) to
obtain $\langle \ln J\rangle$ for odd and even moments yield a new
signal for the transition that coincides with Monte Carlo
simulations. Analysis of high moments yield interesting ``solitonic''
structures that propagate as a function of $p$. Finally, we derive the
exact probability distribution for path sums $J$ up to length $L=64$
for all sign probabilities.
\end{abstract}
\vfill
\eject
\section{Introduction}

Sums of directed paths are present in numerous models of disordered
systems. Polymer configurations in a disordered matrix, dynamics of
interfaces grown by deposition\cite{MedinaHwaKardar} and Feynman path
sums for electron hopping between impurities\cite{NSS,Medina92} are
only a few of the relevant examples. In this paper, we focus on the
latter example involving a model first introduced by Nguyen, Spivak
and Shklovskii (NSS), for interference effects in the strongly
localized regime\cite{NSS}.

In the directed path sign model one studies the sum of all possible
directed paths between two sites on a lattice. On each lattice bond,
one places a random sign with probability $p$. Each directed path
evolved is then computed by multiplying the values of the bonds it
crosses. Finally the sum $J$ of all paths is obtained. The proponents
of the model\cite{NSS} obtained, numerically for small systems, that a
second order transition occurred at $p_c\sim 0.05$ between a phase
with preferential sign (for the path sum $J$), and a phase with no
preferential sign. NSS also offered appealing arguments based on
the behavior of $\delta J/\langle J\rangle$. Presumably, such a
parameter grows exponentially above the transition while it goes to
zero below $p_c$. The physical relevance of this transition lies in
the fact that it may signal the change between Aharonov-Bohm
oscillations of period $hc/e$ and those of $hc/2e$\cite{NSS} in the
context of hopping conduction.

The NSS argument was later contended by Shapir and
Wang\cite{ShapirWang} arguing that correlations between paths implied
that $\delta J/\langle J\rangle$ does not necessarily go to zero for
any $p$. Subsequently, Wang et al\cite{ShapirWangMedinaKardar} used an
exact enumeration scheme to probe the transition for small lattices of
maximum size $L=9$. The work found no evidence of a transition above
negative sign probability $p=0.02$. Such conclusions were supported by
Zhao et al\cite{ZhaoGelfand} on the basis of numerics, for large
square lattices, where it was assured that the transition did not
exist above $p=0.025$ in two dimensions. Nevertheless, the decay of
the order parameter $\Delta P$ as a function of system size was found
to be anomalously slow for finite $p$ (see also
\cite{Medina92}). Therefrom, more recently Spivak, Feng and
Zeng\cite{SpivakFengZeng} discussed numerical results that suggest a
finite jump in the order parameter indicating a first order transition
for the sign problem. The authors also imply that the moments $\langle
J^n\rangle$ increase faster than $n!$ as $n\rightarrow \infty$
indicating there is no unique relation between $\langle J^n\rangle$
and the probability distribution $P(J)$. This is an important point
since the moments, in such a case, may not contain information about
the transition. Finally, in a recent paper by Nguyen and
Gamietea\cite{NguyenGamietea}, a renewed extensive study of the
parameter $\delta J/\langle J\rangle$ proves that, at least according
to such a parameter, no transition exists; only a strong crossover
from logarithmic to exponential behavior is observed.

Besides the numerical approaches, mean field type approximations by
Obukhov\cite{Obukhov} point to a second order transition for
dimensions $d\geq 4$. Furthermore, Derrida and Cook
\cite{DerridaCookSparse} also took up the problem, analytically, using
a sparse matrix approach. They generalized the model to random phases,
which includes random signs as a special case. Their approach is mean
field in nature and results in a phase diagram where the sign
transition is of second order\cite{comment} (see
also\cite{BlumYadin}). Nevertheless, mean field results may not apply
to low dimension due to the importance of path
crossings\cite{ShapirWang}. 

Here we address the following issues: i) What is the order of the sign
transition through scaling analysis of the order parameter proposed,
ii) do moments of the path sums determine the probability distribution
uniquely? and iii) what is the exact behavior of the parameter $\delta
J/\langle J\rangle$ above and below the transition?. A new perspective
will be gained by using hierarchical lattice: Such lattices, while
still amenable to analytical manipulation, include crucial path
correlation effects absent in mean field. 

The paper is organized as follows: Section II discusses the sign model
and describes hierarchical lattices. In section III we perform
detailed Monte Carlo simulations, close to the transition, for systems
of up to size $L=512$. A scaling analysis is performed for the order
parameter $\Delta P=P(J>0)-P(J<0)$ to distinguish between first and
second order transitions. In section IV we study the moments $\langle
J^n\rangle$ exactly, using moment recursion
relations\cite{MedinaKardar93}. We find that moments determine the
distribution uniquely according to Carlemans theorem, and find
possible indications of a phase transition from odd and even moment
extrapolations to $n=0$. In this section we also discuss the high
moment behavior, unveiling interesting structures as a function of the
sign probability $p$. Subsequently, we probe the parameter $\delta
J/\langle J\rangle $ exactly showing, for the first time, its
unambiguous crossover between exponential and logarithmic behavior. In
section V we obtain the exact probability distribution for lattice
sizes $L=16$ and sample the distribution for up to $L=64$ as a
function of $p$. We end with the conclusions and a discussion of the
mapping of the moments to an $n$-body partition function in one
dimension as a continuum model that might aid in explaining the
curious high moment behavior.

\section{The sign model}

Imagine two reference points on a lattice between which one would like
to evolve all possible {\it directed paths} and compute a ``partition
function''
\begin{equation}
\label{partition}
J=\sum_i\Gamma_i,
\end{equation}
where $\Gamma_i$ represents each individual contributing path. By
directed it is meant that paths always propagate in the forward
direction without loops or overhangs. The random medium in which these
paths evolve can be represented by assigning local
weights\cite{KardarOrig} on the bonds or sites that are picked up by
the paths as they wander to their final destination. Such a model has
been used as a paradigm simulating, for example, a coarse-grained
polymer or interface wandering in a random matrix with locally
favorable energy minima\cite{KardarOrig}. The model is interesting
because it yields anomalous lateral wandering and energy exponents for
the interface/polymer as compared to those generated by simple
diffusion, signaling a new disorder induced universality class in
$(1+1)$ dimensions.

Another application, in an entirely different field, is in the context
of Variable Range Hopping\cite{Shklovskii}, a mechanism for conduction
in insulators. In this context, one also needs to sum over Feynman
paths to compute the transition probability, between impurities, of
current bearing electrons. The Feynman paths, in this case, are
directed because they are tunneling paths. Any elongation of the
latter, in the form of loops or overhangs, is exponentially less
probable. For further justification of the model we refer the reader
to the review in reference\cite{Medina92}. NSS studied such tunneling
processes proposing a directed path model where the local weights are
random signs\cite{NSS}. In such a model, the path $\Gamma_i$ is a
product of the signs it picks up en route to the final site. Writing
eq.(\ref{partition}) more explicitly
\begin{equation}
\label{NSSeq}
J=\sum_{\Gamma_i}\prod_i \eta_i,
\end{equation}
where $\eta_i$ is a random sign according to the distribution
$P(\eta)=p\delta(\eta-1)+(1-p)\delta(\eta+1)$. The
probability $p$ in the NSS model emulates the relative abundance of
levels above and below the Fermi energy\cite{NSS}. This model has been
very successful in explaining qualitative and quantitative features of
conduction in the strongly localized regime. In particular, intriguing
interference effects producing a characteristic periodicity of
magnetic field oscillations\cite{reviewShklovskii} and changes in the
localization length due to non-local
effects\cite{Medina92,Milliken}. In spite of the seemingly different
nature of disorder in the NSS model, replica arguments and numerics
have shown that it belongs to the same universality class of directed
polymers with positive weights\cite{Medina92,RouxConiglio}, at least
for $p$ close to $1/2$.

We have taken up the sign model on hierarchical lattices as mentioned
in the introduction. A hierarchical lattice is a recursive structure
built by repeating a chosen motif\cite{BerkerOstlund}. Depending on
the latter motif, one can build integer dimensional objects emulating
an Euclidean lattice. For this work we chose the Berker lattice or
diamond. Such motif (see Fig. \ref{fig0}) has the parameter $b$
corresponding to the number of branches between the initial $i$ and
final $f$ points. The lattice size is related to the recursion order
$m$ as $L=2^{m-1}$ i.e. the number of bonds on any directed path
between $i$ and $f$. The number of bonds on the lattice (or mass) is
given by $M=(2b)^{m-1}$, so that the effective dimension of the
lattice is $d_{eff}=1+\frac{\log b}{\log 2}$. In this work we will use
$b=2$ except if otherwise stated. Qualitative features of critical
behavior of many statistical models are correctly reproduced on such
structures with no unphysical effects. In fact, mapping to
hierarchical lattices is the basis of the Migdal-Kadanoff
renormalization procedure, of frequent use in critical phenomena. As
noted before, an important feature of hierarchical lattices over
either Bethe lattices/mean field approaches is that path intersections
are taken into account. Thus, we expect that the resulting simulations
will be more faithful to low dimensional behavior. In fact, we will
present, in section IV, further evidence of the adequacy of
hierarchical lattices making contact with known recent results on the
sign transition.

\section{Sign phase transition in two dimensions}

In this section, we have undertaken Monte Carlo simulations on
hierarchical lattices to check for scaling properties.  Paradoxically,
scaling has only been discussed once before in connection with the
transition\cite{RouxConiglio} and it is a primary tool to assess its
nature. It will be especially useful to clearly distinguish between
first order and second order transitions.

Hierarchical lattices were generated to $L=512$ or order ten. Averages
were taken over twenty thousand realizations of disorder for a series
of $p$ values between $0$ and $0.5$. As the size of the system
increases more detailed data was collected close to the transition
regime $0.05<p<0.1$. Figure \ref{fig1} shows Monte Carlo data for the
order parameter $\Delta P$ as a function of $p$. A definite plateau at
$\Delta P=1$ develops as $L$ increases for low $p$, signaling a
definite change in the order parameter (positively signed paths
dominate). 

For the proposed order parameter we should expect the scaling form
$\Delta P=f((p-p_c)L^{1/\nu})$. Figure \ref{fig2} shows a good
collapse for the same data as the previous figure. As the order
parameter is always between zero and one, we only need to find $p_c$
and the correlation length exponent $\nu$. For the hypothetical
transition we find the values $p_c=0.071\pm 0.001$ and $\nu=1.85\pm
0.07$ ($1/\nu=0.54$). The latter exponent is very different from that
of percolation on these lattices ($\nu=\ln{2}/\left (
\ln{2}+\ln(3-\sqrt{5})\right )=1.63529..$; so the role of percolation
if any, is not apparent. If the transition were first order the
exponent $1/\nu$ would be the dimensionality of the system
$d$\cite{Kosterlitz}. The non trivial scaling found can also be seen
by taking the derivative of the order parameter and plotting its
maximum as a function of the system size. These criteria rule out a
first order transition.

We have also monitored the evolution of $p_c(L)$ with size. The
specific value of $p_c(L)$ was found from the peak values of the
derivative of the order parameter $\Delta P$. The resulting values are
plotted in Fig. \ref{fig3} where, within error bars, the values of
$1/\nu$ and $p_c(\infty)$ are confirmed. Summarizing, scaling is very
good around $p_c=0.07$ and {\it does not} correspond to the scaling of
a first order transition. Furthermore, there is no sign of a
discontinuity in the order parameter as suggested in
ref.\cite{SpivakFengZeng}. We thus conclude that, on hierarchical
lattices, the transition exists and is second order as mean field
predicts. These conclusions are in agreement with work by Roux and
Coniglio\cite{RouxConiglio} on hierarchical lattices. There, they
analyzed the variable $\alpha_i=(n^+_i-n^-_i)$ where $n^{\pm}_i$ is
the fraction of positive (negative) paths arriving at site $i$, and
they suggest a clear positive $\alpha$ phase. The order of the
transition for hierarchical lattices is not analyzed in detail in
their paper. Nevertheless, they noted undue emphasis of hierarchical
lattices on the $\alpha=0$ result, and the possible impact of this on
the scaling properties of various quantities. We will come back to
such observations, briefly, in section V.

\section{Moment recursion relations}

A statistic we can probe exactly on hierarchical lattices are the
moments of the probability distribution. This is possible because of
recursion relations derived by Cook and Derrida\cite{DerridaCook} and
generalized to arbitrary moment and hierarchical order (system size)
by Medina and Kardar\cite{MedinaKardar93}. The recursion relation for
$b=2$ is the following
\begin{equation}
\label{recursion}
\langle J_{m+1}^n\rangle=\sum_{s=0}^{n}\frac{n!}{(n-s)!s!}[\langle
J_{m}^s\rangle]^2[\langle J_{m}^{n-s}\rangle]^2,
\end{equation}
where $n$ is the moment number and $m$ is the hierarchical lattice
order. This expression is readily generalized to other integer $b$ by
changing the binomial factor to a multinomial and including the
additional branches. Hence, one can emulate higher dimensional
networks. The simple form of this recursion permits, given the local
moments at order one, to compute moments to any given lattice
size. Appropriate programming of the recursion relations, with
arbitrary precision computations, is linear in time with lattice
order.

The behavior of the moments for the sign model is extremely rich as we
shall see in the following. As found in ref. \cite{MedinaKardar93},
after a few hierarchical orders, the values $ln(\langle
J_m^n\rangle)/L$ converge rapidly to a limiting form as a function of
$n$. Such limiting form is important because it also signals the
convergence to a unique limiting distribution, at least, if moments do
not grow faster than $n!$\cite{Feller}. The asymptotic form of the
moments can be obtained for $p=0$\cite{MedinaKardar93,DerridaCook},
\begin{equation}
\label{pequalzero}
{\ln\langle J^n\rangle\over L} = n\left (1-\frac{1}{L}\right )ln 2,
\end{equation}
that is, moments grow exponentially with $n$ for $p=0$. Nevertheless,
for $0<p< 1/2$, lower moments grow slightly faster than exponential
($\exp(n^{\alpha})$ with $1<\alpha<2$), gradually converging to
exponential growth for larger moments. The latter implies, according
to the condition
\begin{equation}
\label{Carleman}
\sum_{n=0}^{\infty}\langle J^{2n}\rangle^{-1/2n}=\infty,
\end{equation}
that the moments determine the distribution uniquely. There are
various forms of such a theorem, but the above is the strongest
version due to Carleman\cite{Feller}. If one substitutes above,
$\langle J^{2n}\rangle\sim \exp(2n)$, -our asymptotic result- the
criterion is satisfied. Even if $\langle J^{2n}\rangle$ grows slightly
faster i.e. $\exp(2n\ln 2n)\sim (2n)!$ the above sum diverges because
$\sum_n 1/n=\infty$. Any faster growth would violate
eq.(\ref{Carleman}), factorial growth being the borderline case. That
the moments $\langle J^n\rangle$ satisfy eq.(\ref{Carleman}) is one of
our central results. In Fig. \ref{fig4} we show a sequence of moments
as a function of the moment number $n$. The different curves, starting
from below, represent hierarchical orders one thru nine (sizes $L=2$
thru $L=256$). One readily notes convergence to a definite law. The
inset shows a comparison between the growth of $n!$ and that of
moments for the particular case of $p=0.1$. The asymptotic behavior is
already reached at $L=128$, larger sizes falling on the same curve.

For values close to $p=1/2$, the moment sequence has a characteristic
sawtooth shape, where even moments are at the crests and the odd at
the troughs. Such structure is not a finite size effect. We have
checked this for up to $L=2^{20}$ on the hierarchical lattice. As
$p\rightarrow 1/2$ all the odd moments go to zero while the even
remain finite as expected. On the other hand, as $p$ is reduced the
sawtooth disappears, first for the higher moments and then for the
lower. In this respect there appears to be a phase transition for each
moment at different values of $p$, in a way reminiscent of that
discussed by Cook and Derrida\cite{DerridaCook} (in their case as a
function of `temperature'). The transition for the first two moments
occurs close to $p=0.075$ which is close to that found from Monte
Carlo simulations in the previous section. On this basis it is
plausible that the disappearance of sawtooth shape is related to the
transition.

Figure \ref{fig5} shows a set of curves for $d\ln(\langle
J_m^n\rangle/L) /dn$, and various values of $p$, $L=2^{17}$ and up to
$n=100$. The last six orders of the hierarchical lattice collapse
on the same curve indicating we have achieved asymptotics. For the
highest value of $p$ one notes the sawtooth behavior, while it
disappears for all moments below $p=0.1$. Nevertheless, additional
structure is observed at moments beyond $n=40$ for $p=0.075$ and
$p=0.1$, where a shoulder develops and moves towards larger $n$ values
as $p$ increases, undeformed, in a solitonic manner. Although the
analysis of these structures is out of the scope of this paper, it is
interesting to analyze it in the light of a mapping to a one
dimensional many body problem\cite{Zhang}. In such a mapping the
moment number corresponds to the number of particles interacting like
charges on contact. Thus we speculate that the shoulders could be
related to sudden changes in the character of the ground state as the
particle number (moment number) increases. We will discuss this in
more detail in the final section.

For even smaller $p$ values the curve starts to resemble the well
known $p=0$ limit given by eq.(\ref{pequalzero}), and depicted as a
flat line at $\ln 2$ in Fig. \ref{fig5}.  From the figure one can
graphically identify the value of $\langle \ln J\rangle$ as a function
of $p$ using the relation $d\ln\langle J_m^n\rangle/dn|_{n\rightarrow
0}=\langle \ln J\rangle$. The quantity $\langle \ln J\rangle$ is a
``free energy'' that may reflect the sign transition. We have followed
the value at intercept mentioned before as a function of $p$ below
$p=0.2$. When the moments ``zigzag'' there are two possible
extrapolations, while below the assumed transition the moments lead to
a single prediction of the free energy. The results are depicted in
Fig. \ref{fig6}, where the curves merge around $p_c=0.07$ within the
error of the extrapolation procedure. Such a value coincides with our
Monte Carlo prediction.

One can validate the relevance of hierarchical lattices by checking
the exact computation of the variable $\delta J/\langle J\rangle$ with
$\delta J=\sqrt{\langle J^2\rangle -\langle J\rangle^2}$. Such a
quantity has been discussed extensively in previous
work\cite{NSS,ShapirWang,NguyenGamietea,ShklovskiiSpivak85}. As
mentioned before, $\delta J/\langle J\rangle$ was initially suggested
as a candidate for a kind of order parameter that diverged
exponentially above the transition and went to zero
below. Observations by Shapir and Wang\cite{ShapirWang} showed,
nevertheless, that path correlations (crossings) invalidated the
vanishing of the parameter for any value of $p$. It has been argued
that for small $p$ there is a crossover from exponential growth (for
$p>p_c$) to logarithmic growth (for
$p<p_c$)\cite{NguyenGamietea,ShklovskiiSpivak85}. Shapir and Wang, on
the other hand found a change from $\exp[|log(1-2p)|2a\sqrt{L}]$ for
$p<p_c$ to $\exp[|log(2(1-2p)^2)|2L]$ for $p>p_c$. Yet, they observe
that the former result is incorrect because partial overlaps of pairs
of walks should be accounted for.

Simulations on regular lattices to date can only do very poorly in
proving the surmised logarithmic behavior below $p_c$. Here we have
computed $\delta J/\langle J\rangle$ to sizes $L=2^{20}$ for various
$p$ values in a few CPU minutes. We have found a clear confirmation of
logarithmic to exponential crossover as $p$ increases. Figures
\ref{fig7} and \ref{fig8} show $\delta J/\langle J\rangle$ and its
derivative as a function of $L$ respectively. The scales used permit
rapid identification of the corresponding behavior. It should be noted
that, on euclidean lattices, it is reported the behavior reported is
$\delta J/\langle J\rangle\propto (\ln L)^\mu$, where $\mu\sim 1$ but
depends weakly on $p$.

On hierarchical lattices we can also demonstrate analytically
that there is no transition in the variable $\delta J/\langle
J\rangle$. Following Cook and Derrida\cite{DerridaCook},
eq.(\ref{recursion}), for the first two moments, can be written for
general $b$ as
\begin{eqnarray}
\label{JRecur}
\langle J_{m+1}\rangle & = & b\langle J_{m}\rangle^2,\nonumber\\
\langle J^2_{m+1}\rangle & = & b\langle J^2_m\rangle^2+b(b-1)\langle
J_m\rangle^4.
\end{eqnarray}
Now, after defining $j_2(m)=\langle J_m\rangle^2/\langle J_m^2\rangle$
one can write a recursion relation for $\delta J/\langle J\rangle$ as
\begin{equation}
\label{RecurDeltaJoverJ}
\left (\frac{\delta J_{m+1}}{\langle J_{m+1}\rangle}\right
)^2=\frac{1}{b}\left [\frac{1-j_2^2(m)}{j_2^2(m)}\right ].
\end{equation}
It is simple to determine that $j_2$ has in general three fixed
points, $j_2=0,1,1/(b-1)$. For $b>2$ ($d_{eff}>2$) a critical fixed
point arises and $\delta J/\langle J\rangle$ exhibits a phase
transition as NSS proposed. On the other hand, for $b=2$ there are
only two trivial fixed points; $j_2=1$ is unstable and $j_2=0$ is
stable, indicating that $\delta J/\langle J\rangle$ always diverges as
found above. Values of $j_2$ close to one correspond to $p\rightarrow
0$, while $j_2$ close to zero correspond to $p\rightarrow 1/2$.
Analyzing the behavior of the recursion for $j_2$ near the $j_2=0$
fixed point one can derive from Eq.\ref{JRecur} that $\delta J/\langle
J\rangle \sim 1/2\exp(L(|\ln j_2(0)|+(1/2)j^2_2(0)))$. The behavior
close to $j_2=1$, which should be logarithmic, is also verified
(numerically) although we have not arrived at a simple closed
expression.  In summary, hierarchical lattices provide similar results
to those expected on euclidean lattices thus seeming a good testing
ground for the sign transition.

As a final word; we have computed higher order cummulants of $J$
finding no features of special interest related to the transition. The
only result worth mentioning is that $\ln (C_j)^{1/j}/L=\ln2$ for
$p=0$, where $C_j$ is the $j$th cummulant of $J$. In what follows we
will take advantage of the special structure of hierarchical lattices
to compute the full probability distribution for $J$.

\section{Probability distribution for $J$}

Monte Carlo sampling of the distribution of $J$ is handicapped by the
models distribution broadness. For such reasons Wang et
al\cite{ShapirWangMedinaKardar} undertook an exact enumeration study
to probe the NSS order parameter $\Delta P=P(J>0)-P(J<0)$. Because of
the high computer demand of exact enumeration, they could only
access sizes of $L=10$ for all $p$. Here we use a scheme, on
hierarchical lattices, permitting access to $L=16$ exactly for all $p$
and sampling of the distribution for $L=64$. The procedure is as
follows: As a hierarchical lattice is built recursively following a
chosen motif, one can write the following recursion relation for the
probability distribution.
\begin{equation}
\label{ProbRec}
P_{l+1}(J)=\prod_{i=1}^4\int_{-\infty}^{\infty}P_l(\eta_i)\delta
(J-\eta_1\eta_2-\eta_3\eta_4)d\eta_i,
\end{equation}
where $\eta_{1,2}$ and $\eta_{3,4}$ denote contiguous elements on
separate branches of the hierarchical
lattice. $P_1=p\delta(\eta-1)+(1-p)\delta(\eta+1)$, where $p$ is the
sign probability discussed in previous sections. The number of
possible outcomes for $J$ or number of different paths goes as
$2^{2^{m-1}-1}$ ($32768$ for $L=16, m=5$, and $2147483648$ for $L=32,
m=6$). This growth is extremely fast, although many $J$ values will be
degenerate for any particular disorder realization. Note that while
$L=16$ is easily accessible, going an order further, puts the
calculation out of reach, no intermediate sizes being available on
hierarchical lattices. For $L=32$ we have resorted to a
coarse-graining procedure in the following manner: the exact results
for $L=16$ involve $175$ terms which we cannot evolve exactly to the
next order. Nevertheless, we can make a coarse-grained distribution by
averaging $J$ occurrences in groups of $7$ to obtain $25$ different
values. One can then go to up to $L=64$ by repeating this
procedure. Beyond such a size, the coarse graining procedure does not
incorporate sufficient detail to see anymore changes in the
distribution, so within our resolution we have achieved its limit
form.

Fig. \ref{fig9}a shows the probability distribution for $L=16$ for
significant values of $p$. The probability distribution is
astonishingly complex, even for small sizes, revealing rich
interferences in the paths sums $J$. Note that it would be hopeless to
sample the distribution $P(J)$ using Monte Carlo as there are sixty to
one hundred and thirty orders of magnitude of
probability. Fig. \ref{fig9}b shows different $p$ values for a sample
$L=32$ as a function of $p$ using the coarse-graining procedure
described above. As expected the distribution is symmetric for $p=0.5$
and gains asymmetry ($\Delta P\neq 0$) as $p$ moves towards zero. Note
that $P(J)$ falls slower than exponential, on average, about the
peak value. It is notable the speed with which the distribution
appears symmetric beyond $p=0.1$. This feature is understood in the
`zigzag' behavior of the moments, where odd moments are much smaller
than even moments and their separation increases exponentially as $p$
is increased. 

Having the information of the exact distribution one is also able to
obtain the exact order parameter $\Delta P$ introduced in section
III. No qualitative differences were found with curves reported in
figures 1 and 2 at least to sizes $L=32$, so sampling of $\Delta P$,
involved in Monte Carlo, seems to be good enough to draw the
conclusions about the transition (see section III).

In Fig. \ref{fig10} we have depicted the distribution for $L=64$ and
$p=0.5$ without joining the points for the probability amplitude (as
was done in Fig. \ref{fig9}). A fractal structure is apparent; The
whole distribution, in the shape of an approximate triangle, is built
from, scaled, identical triangular structures up to the resolution
achieved by the coarse-graining procedure. Similar complexity is
expected for the sign problem on Euclidean lattices.

An interesting final point to make in this section is that, in view of
the unique relation between distribution and moments (see section IV),
it is possible to use known inversion formulas\cite{Feller}. In this
way one could derive the limiting distribution exactly to any order
desired.

\section{Summary and Discussion}

We have provided evidence for the existence of a phase transition for
the directed path sign model on hierarchical lattices. Nontrivial
finite size scaling of the order parameter close to the transition,
points to a second order phase transition as found from mean field
type approaches. From numerical computations, the threshold on diamond
hierarchical lattices is $p_c=0.071\pm 0.001$ and the correlation
length exponent is $\nu=1.85\pm 0.07$. The latter exponent is very
different from that of percolation on the same lattice $\nu=1.635..$.

The study of exact moment recursion relations for $\langle J^n\rangle$
led us to the definitive conclusion that the moments uniquely
determine the probability distribution, according to Carleman's
theorem\cite{Feller}. Using extrapolations of the derivative of
integer moments ($d\langle J\rangle^n/dn$) to $n=0$, we were able to
find a ``free energy'' $\langle \ln J \rangle$. Such a free energy
splits into two possible extrapolations (from even and odd moments) as
one goes through the transition point by increasing $p$. The latter
transition point coincides with that found in Monte Carlo
simulations of the sign transition. We have not completely interpreted
this connection in the present paper. Furthermore, we studied the high
moments of the partition function $J$ below the transition, and found
a very interesting non-monotonic behavior including step structures
that propagate on the moment number axis, as $p$ changes. 

Using the fact that moments can be computed exactly we studied the
celebrated ratio $\delta J/\langle J\rangle$ proposed by NSS. We have
shown, analytically that indeed in $d_{eff}=2$ the ratio does not show
a transition as suspected
numerically\cite{ShapirWang,NguyenGamietea,ShklovskiiSpivak85} on
regular lattices. Furthermore, we have shown that hierarchical
lattices exhibit the same logarithmic to exponential crossover for
$\delta J/\langle J\rangle$ surmised in references
\cite{NguyenGamietea} and \cite{ShklovskiiSpivak85}.

Finally, we studied a recursion relation for the full probability
distribution for $J$, finding an extremely complex structure even for
systems as small as $L=16$. Previous remarks by Roux and
Coniglio\cite{RouxConiglio} of anomalous accumulation of probability
at $J=0$ are confirmed. Nevertheless, their claim that the
hierarchical lattice becomes essentially one dimensional for large
$L$, and thus, the probability distribution should approach a Gaussian
is not borne out from our results. One obvious difference is that for
a Gaussian all cumulants, larger than two, should be zero which is in
disagreement with exact results of section IV. No evident signal of
the transition, beyond that already inferred from the order parameter
$\Delta P$, is found from the full probability distribution.

Medina and Kardar\cite{Medina92} have studied the moments for the sign
problem interpreting them as partition functions for $n$-body one
dimensional Hamiltonians with contact interactions. Most of the focus,
however, has been on the low $n$ behavior that yields cummulants of
$\ln J$. Nevertheless, it would be interesting to interpret the
findings of this paper, regarding high moments, in the light of a many
body-theory. A previous effort by Zhang\cite{Zhang} focused on the
Hartree-Fock approximation valid only for a large number of particles
(higher moments). In Zhang's approach the sign model was equivalent to
finding the ground state of the many body Hamiltonian
\begin{equation}
\label{ZhangEq}
\left (-\sum_{i=1}^{2n}\partial_i^2+\sum_{i>j}e_i
e_j\delta(x_i-x_j)\right )\Psi(x_1...x_n)=E_0(n)\Psi(x_1...x_n)
\end{equation}
where $e_i$ is a charge that acts via contact interaction of the
$i_{th}$ particle: $e_i=1$ for $1\leq i \leq n$ and $e_j=-1$ for
$n\leq i\leq 2n$. Zhang's approach yields $E_0\propto n^2$. Our
findings predict, from the relation $\ln \langle J^n\rangle/L=E_0$,
$E_0=\gamma n$ for large $n$ (see also \cite{Medina92}), where
$\gamma$ increases as $p\rightarrow 0$. For lower $n$ the behavior is
non-trivial and is certainly not represented a simple powerlaw of
$n$. Therefore, Zhangs results represent some kind of intermediate
regime. A more detailed solution of eq. (\ref{ZhangEq}) might yield
the `solitonic' patterns reported here (see section IV) which are not
well understood.  As speculated in section IV, the ground state formed
by particles with attractive and repulsive interactions might change,
suddenly, at critical particle numbers generating discontinuities in
the derivative of $\ln \langle J^n\rangle$. More work is needed in
this direction.

The highly non-monotonic behavior displayed by the moments calls for
caution regarding the regime of validity of moments dependencies on
the moment number $n$ reported in the
literature\cite{Medina92,Zhang}. Claims of a non-unique relation
between moments and the probability distribution\cite{SpivakFengZeng}
were based on expressions only valid in the $n\rightarrow 0$ limit,
which is clearly unrelated to the constraints of Carlemans
theorem\cite{MedinaUnpub}. Obviously, the conclusions of this paper
are only valid in the measure to which hierarchical lattices agree
with continuum results. For a discussion of the latter point see
reference\cite{MedinaKardar93}.

\acknowledgements EM thanks G. Urbina and A. Hasmy for careful reading
of the manuscript and R. Paredes for a valuable suggestion. EGA thanks
R. Rangel for useful discussions. This work was supported by CONICIT
through grant S1-97000368, a grant from the Research Fund of UNEXPO
and the POLAR Foundation.

\begin{figure}
\caption{Hierarchical lattices are built by repeating a chosen motif;
each bond turns into a diamond recursively. The figure shows successive
iterations of the lattice and the corresponding length $L$ between end
points $i$ and $f$. Examples of a directed path at each order are
indicated by contiguous arrows.}
\label{fig0}
\end{figure}

\begin{figure}
\caption{The figure shows the order parameter $\Delta P=P(J>0)-P(J<0)$
as a function of the sign probability $p$ for system sizes
indicated. Averages were performed over more than $20000$ realizations
of randomness. Note the formation of a plateau at $\Delta P=1$ for
small $p$.}
\label{fig1}
\end{figure}

\begin{figure}
\caption{Same data as in Fig. \ref{fig1} after collapsing the curves for
different system sizes. The appropriate choices for $p_c$ and $\nu$,
the transition threshold and the correlation length exponent, are
indicated.}
\label{fig2}
\end{figure}

\begin{figure}
\caption{The figure shows the value of $p_c(L)$, evaluated from the peaks 
of the derivative of the order parameter, as a function of
$(1/L)^{1/\nu}$. The last five sizes from $L=32$ to $L=512$ have been
fitted by least squares to yield the asymptotic value
$p_c(\infty)=0.072$ indicated.}
\label{fig3}
\end{figure}

\begin{figure}
\caption{The moments $\ln \langle J^n\rangle/L$ as a function 
of the moment number $n$ for the lattice sizes indicated. The figure
shows the rapid convergence to an asymptotic result. In the inset, we
show that while the initial moments grow faster than exponential they
nevertheless grow slower than $n!$, so there is a unique relation
between moments and probability distribution.}
\label{fig4}
\end{figure}

\begin{figure}
\caption{The derivative of $\ln \langle J^n\rangle/L$ as a function of
the moment number, for size $L=2^{15}$, (last six orders collapse onto
the same curves.) As $p$ decreases the curves approach the asymptotic
value $\ln 2$. Note the change from the sawtooth behavior above
$p=0.07$ to collinearity. Shoulder features, developing at higher $n$,
move almost undeformed in the positive $n$ direction as $p$
increases.}
\label{fig5}
\end{figure}

\begin{figure}
\caption{The ``free energy'' $\langle \ln J\rangle/L$, as discussed in
the text, as a function of $p$. Note that the curves extrapolated from
even and odd moments merge around the threshold for the transition
obtained from Monte Carlo.}
\label{fig6}
\end{figure}

\begin{figure}
\caption{The figure shows the behavior of $\delta J/\langle J\rangle$ as a 
function of $L$ for the $p$ values indicated. As the plot is semi-log
the exponential behavior above $p=0.03$ is evident.}
\label{fig7}
\end{figure}

\begin{figure}
\caption{The figure shows the derivative of the data in
Fig. \ref{fig7}. Here the logarithmic behavior of $\delta J/\langle
J\rangle$ is evident. The dotted line is a guide for the eye for $1/L$
behavior.}
\label{fig8}
\end{figure}

\begin{figure}
\caption{a) The exact probability distribution for $L=16$ at the $p$
values indicated. Note the self-affine structure; the central peak is
repeated at subsidiary local maxima. b) The figure shows the
coarse-grained distribution for $L=32$.}
\label{fig9}
\end{figure}

\begin{figure}
\caption{The probability distribution for $L=64$ and $p=0.5$ using the
coarse-grained distribution for $L=32$. To emphasize the self-similar
structure we have not to joined the points in the graph as in
Fig. \ref{fig9}.}
\label{fig10} 
\end{figure} 


\begin{references} 

\bibitem{MedinaHwaKardar} E. Medina, T. Hwa, M. Kardar, and Y.-C
Zhang, Phys. Rev. A {\bf 39}, 3053 (1989).

\bibitem{NSS} V. L. Nguyen, B. Z. Spivak, and B. I. Shklovskii, Pis'ma
Zh. Eksp. Teor. Fiz/ {\bf 41}, 35 (1985) [JETP Lett. {\bf 41}, 42
(1986); Zh. Eksp. Teor. Fiz. {\bf 89}, 11 (1985) [Sov. Phys.-JETP {\bf
62}, 1021 (1985)].

\bibitem{Medina92} E. Medina, and M. Kardar, Phys. Rev. B {\bf 46},
9984 (1992).

\bibitem{ShapirWang} Y. Shapir, and X. R. Wang, Europhys. Lett. {\bf
4}, 1165 (1987).

\bibitem{ShapirWangMedinaKardar} X. R. Wang, Y. Shapir, E. Medina, and
M. Kardar, Phys. Rev. B {\bf 42}, 4559 (1990).

\bibitem{ZhaoGelfand} H. L. Zhao, B. Spivak, M. Gelfand, and S. Feng,
Phys. Rev. B {\bf 44}, 10760 (1991).

\bibitem{SpivakFengZeng} B. Spivak, S. Feng, F. Zeng, JETP Lett. {\bf
64}, 312 (1996).

\bibitem{NguyenGamietea} V. L. Nguyen, and A. D. Gamietea,
Phys. Rev. B {\bf 53}, 7932 (1996).

\bibitem{Obukhov} S. Obukhov, in {\it Hopping Conduction in
Semiconductors} ref.\cite{reviewShklovskii} p.338.

\bibitem{DerridaCookSparse} J. Cook, and B. Derrida,
J. Stat. Phys. {\bf 61}, 961 (1990).

\bibitem{comment} In ref.\cite{DerridaCookSparse} Cook and Derrida add
a temperature parameter. The limit $T\rightarrow \infty$ corresponds
to the pure sign model we treat here.

\bibitem{BlumYadin} Y. Y. Goldschmidt, and T. Blum, J. Phys. I {\bf
2}, 1607 (1992).

\bibitem{MedinaKardar93} E. Medina, and M. Kardar,
Jour. Stat. Phys. {\bf 71}, 967 (1993).

\bibitem{KardarOrig} M. Kardar, Y.-C. Zhang, Phys. Rev. Lett. {\bf
58}, 2087 (1987).

\bibitem{Shklovskii} B. I. Shklovskii, and A. L. Efros, {\it Electronic
Properties of Doped Semiconductors} (Springer-Verlag, Berlin, 1984).

\bibitem{reviewShklovskii} B. I. Shklovskii and B. Z. Spivak, in {\it
Hopping Conduction in Semiconductors}, edited by M. Pollak and
B. I. Shklovskii (North-Holland, Amsterdam, 1990).

\bibitem{Milliken} F. P. Milliken, and Z. Ovadyahu,
Phys. Rev. Lett. {\bf 65}, 911 (1990).

\bibitem{RouxConiglio} S. Roux, and A. Coniglio, J. Phys. A {\bf 27},
5467 (1994).

\bibitem{BerkerOstlund} A. N. Berker, and S. Ostlund, J. Phys. C {\bf
12}, 4961 (1979).

\bibitem{Kosterlitz} J. Lee, and J. M. Kosterlitz, Phys. Rev. B {\bf
43}, 3265 (1991).

\bibitem{DerridaCook} J. Cook, and B. Derrida, J. Stat. Phys. {\bf
57}, 89 (1989).

\bibitem{Feller}W. Feller, {\it An Introduction to Probability
Theory and Its Applications} (Wiley, New York, 1971).

\bibitem{Zhang} Y.-C. Zhang, Jour. Stat. Phys. {\bf 57}, 1123 (1989).

\bibitem{ShklovskiiSpivak85} B. I. Shklovskii, and B. Z. Spivak,
J. Stat. Phys. {\bf 38}, 267 (1985).

\bibitem{MedinaUnpub} E. Medina, unpublished.


\end{references}
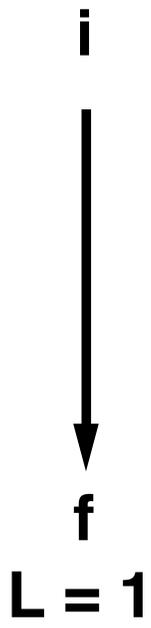
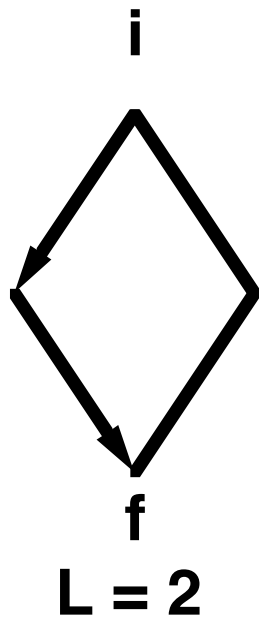
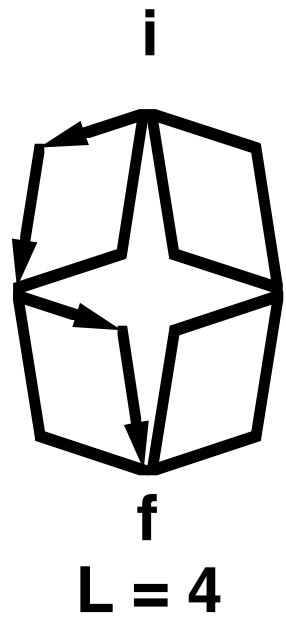
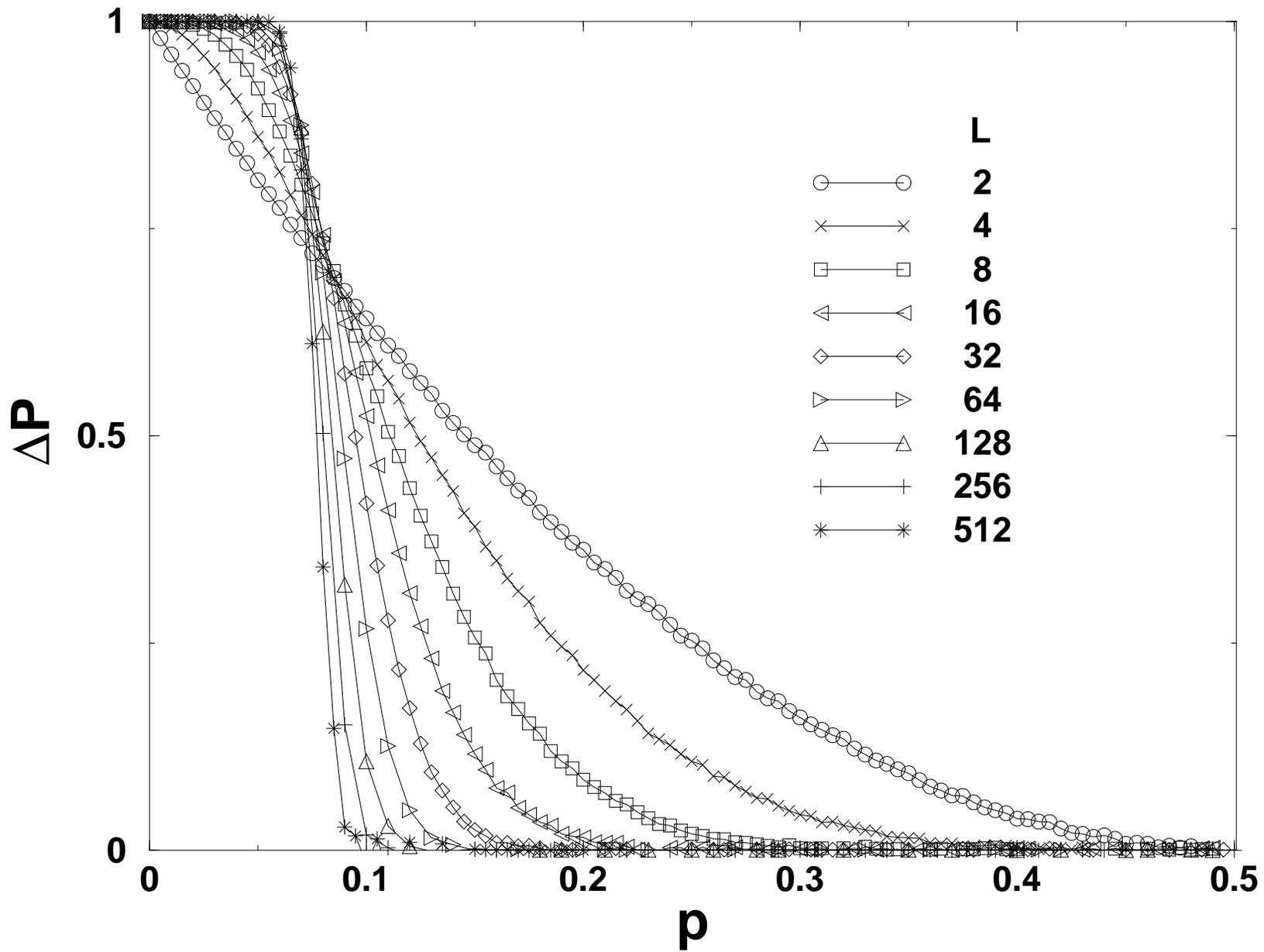
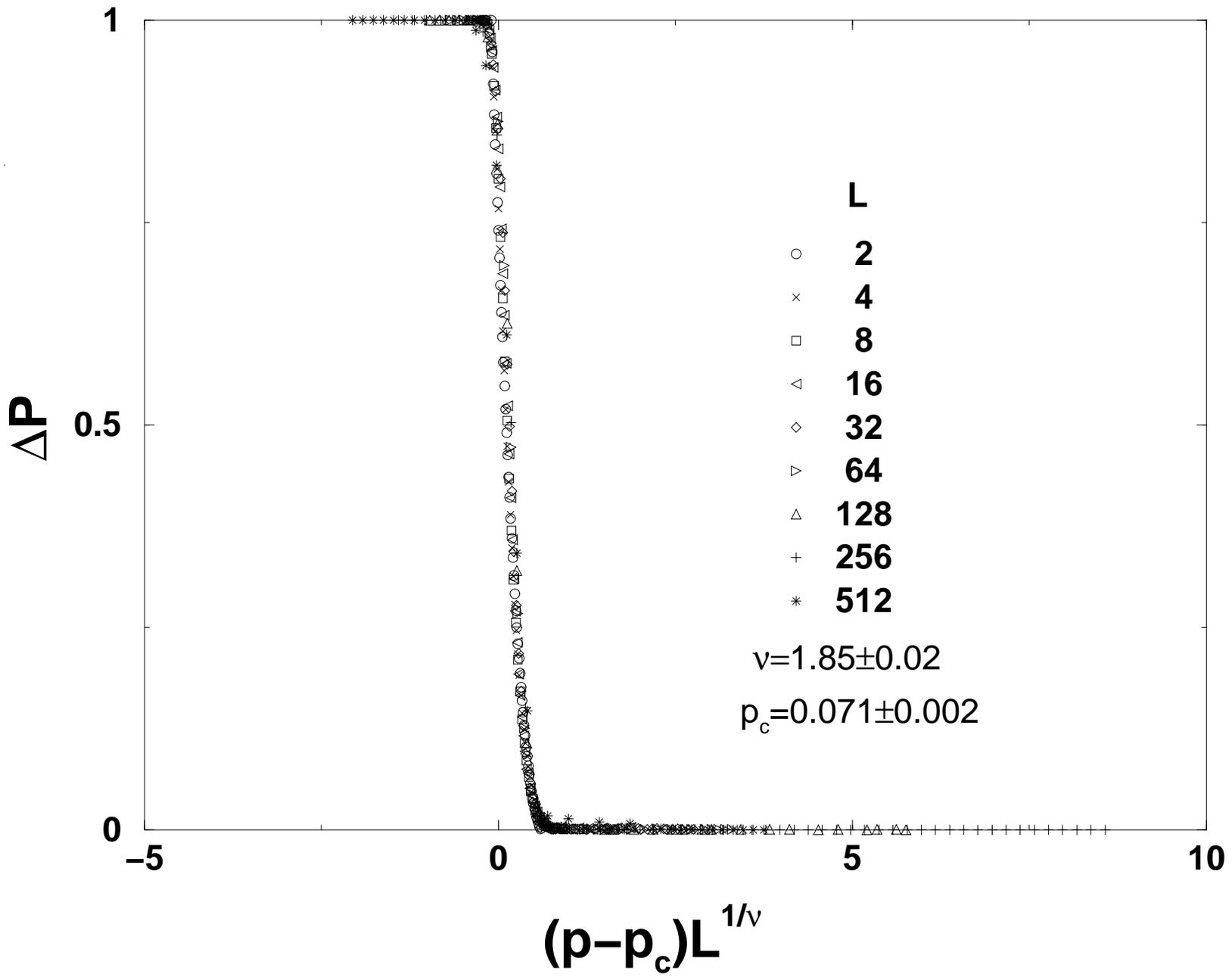
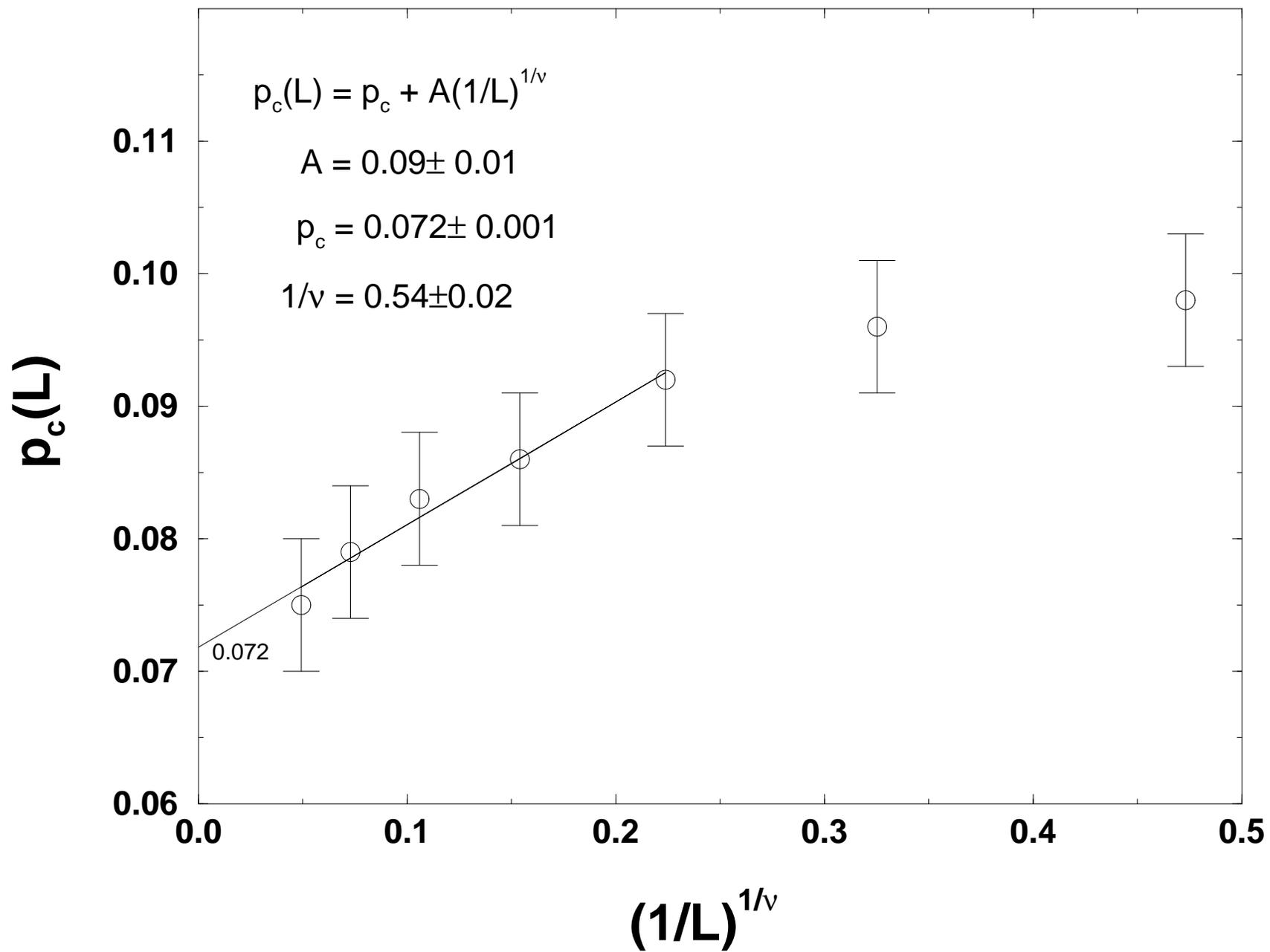
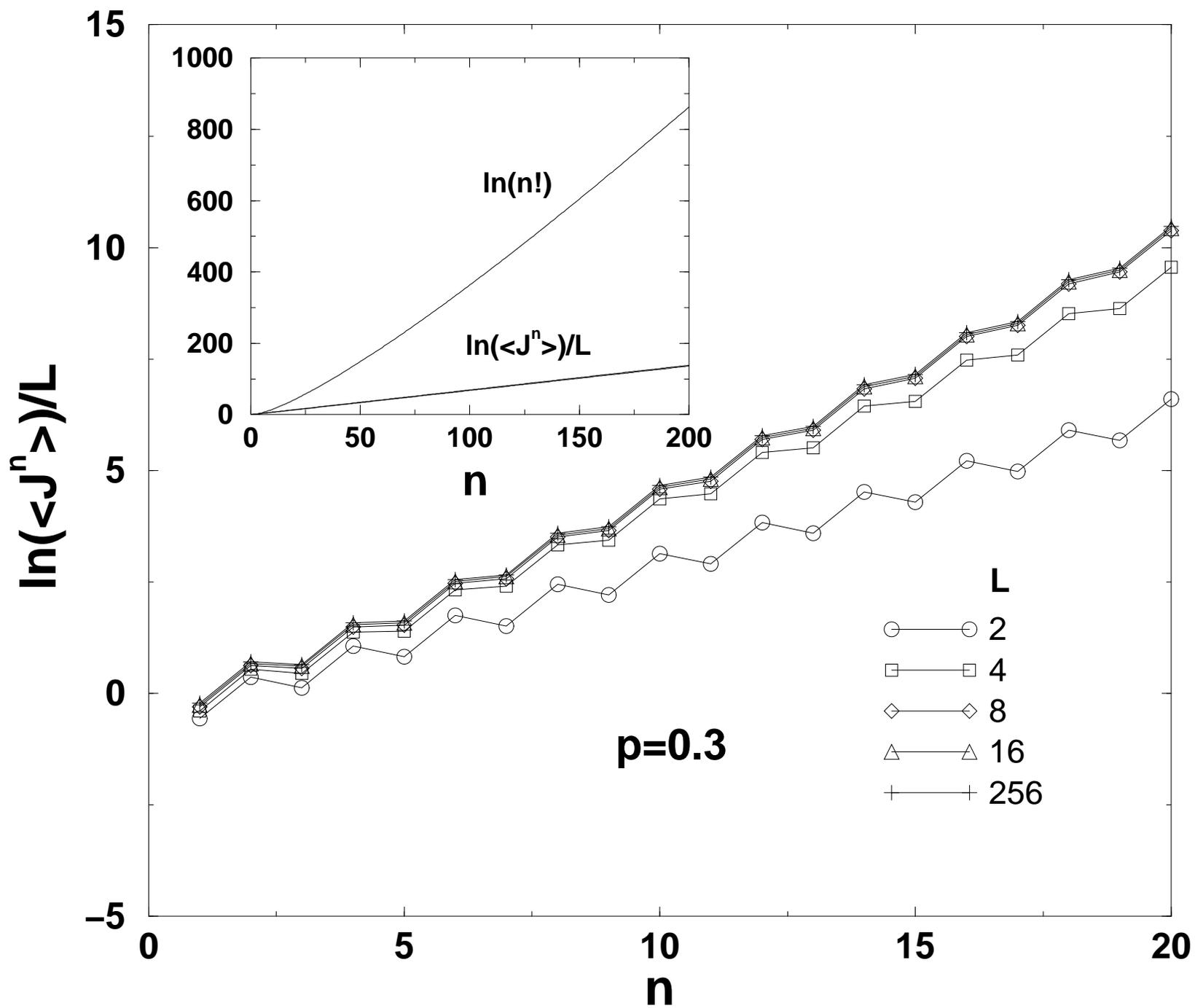
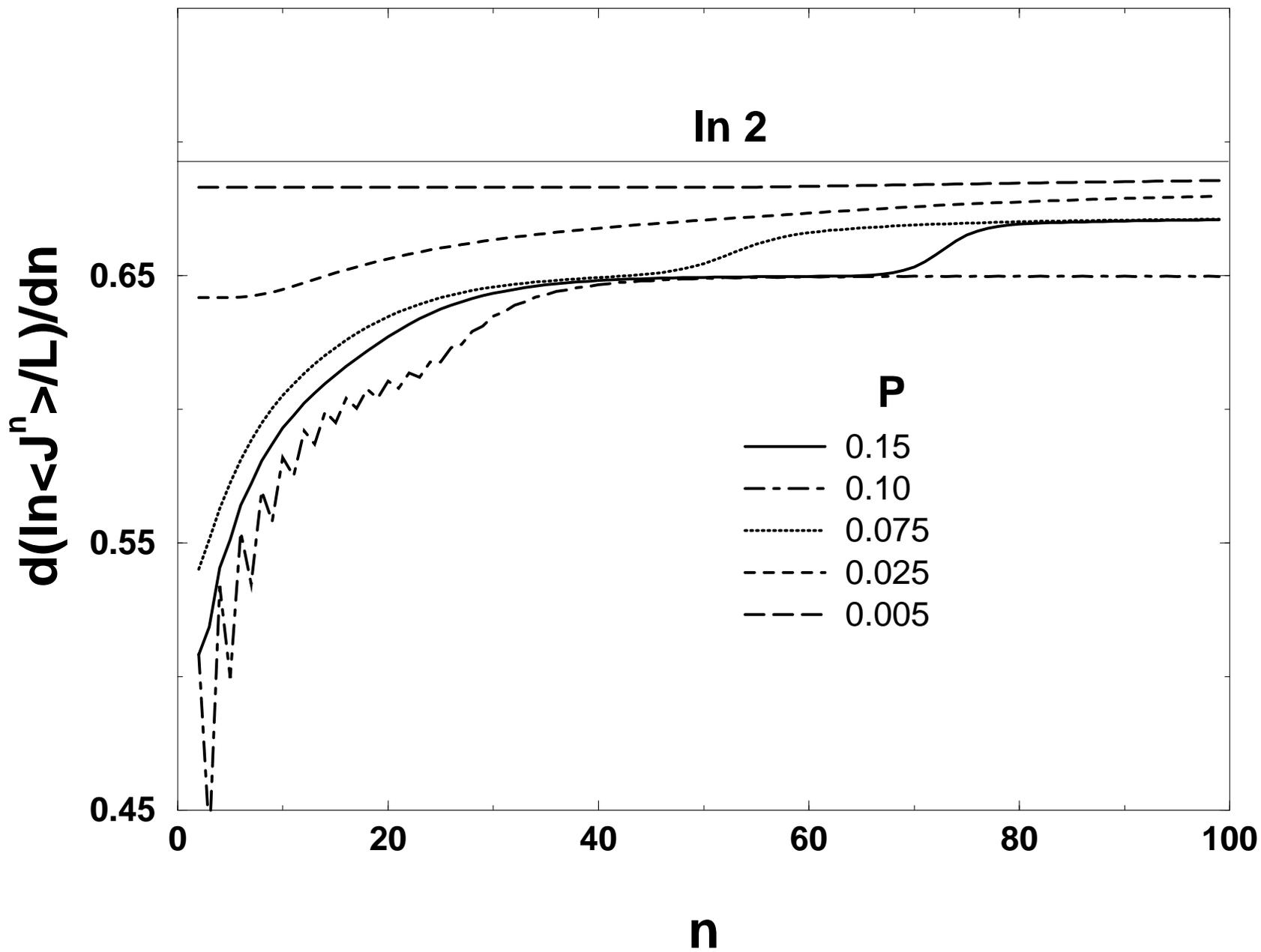
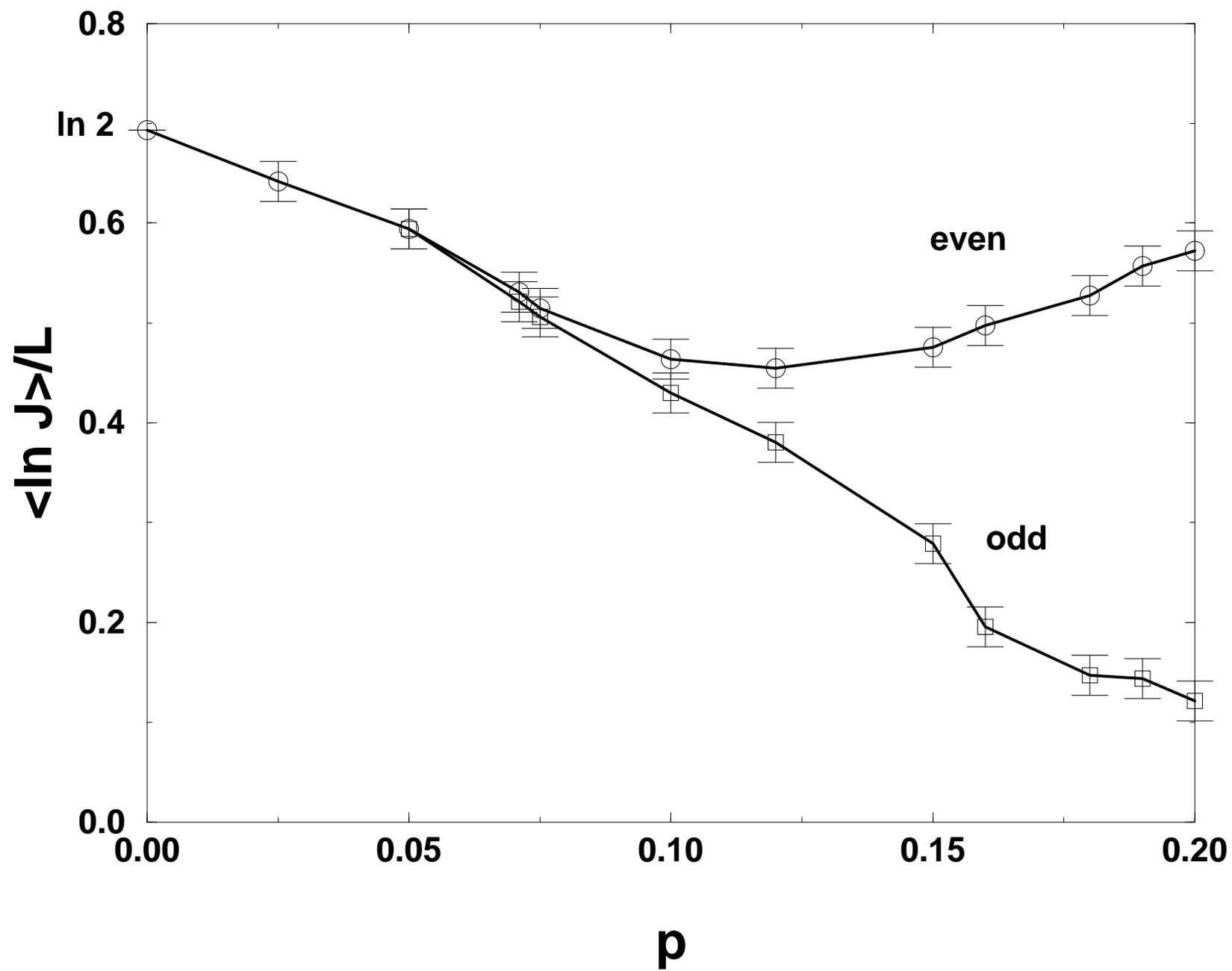
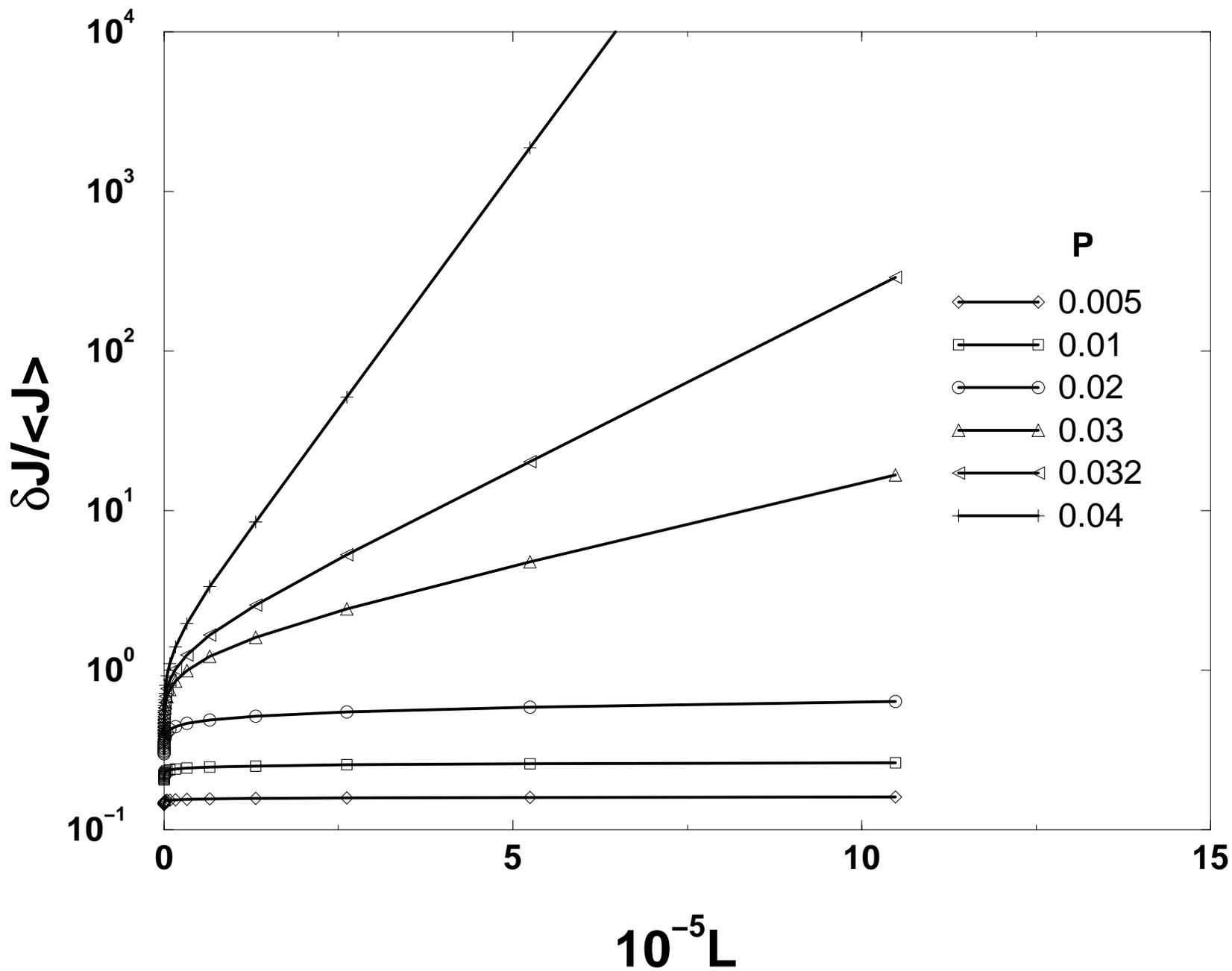
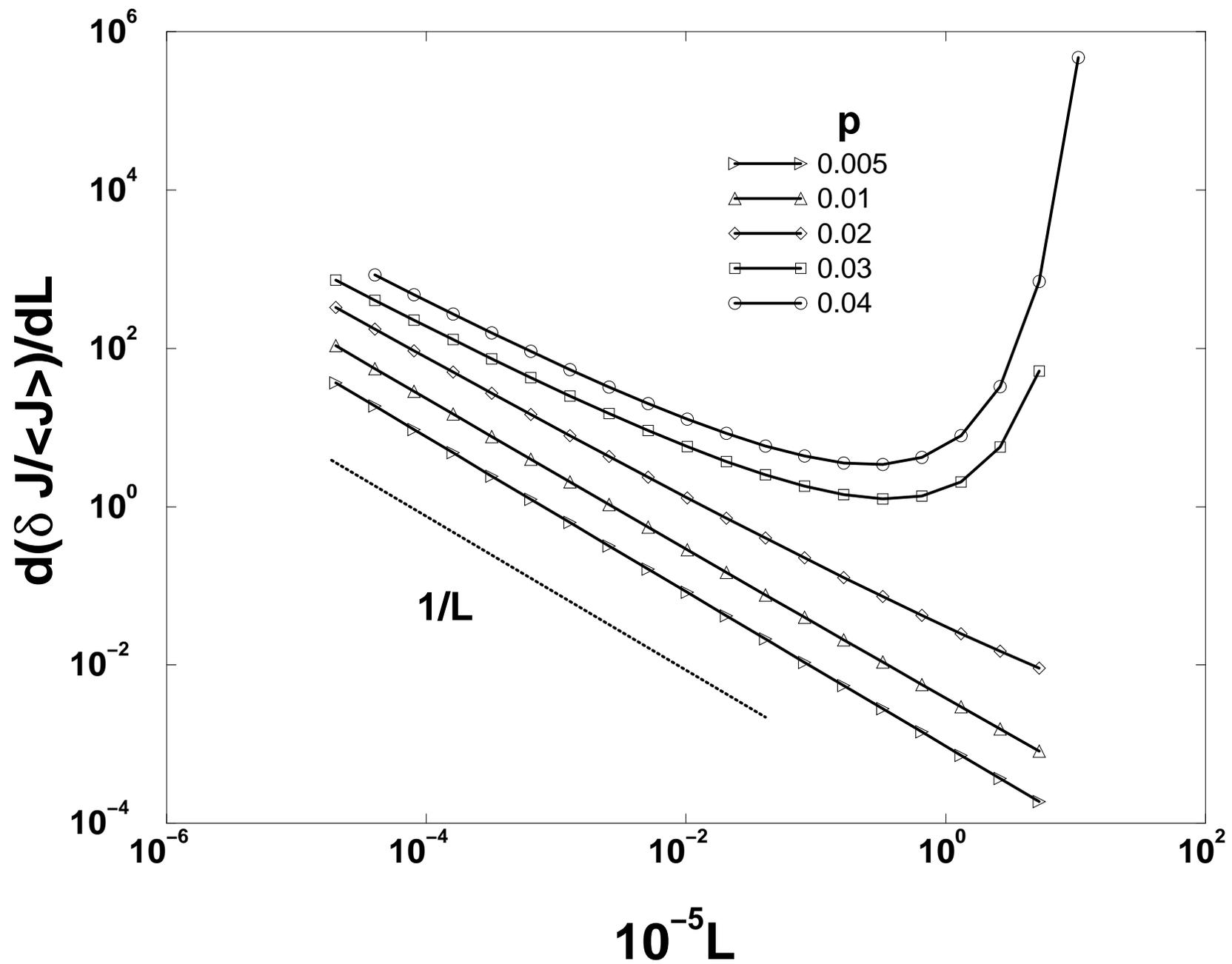
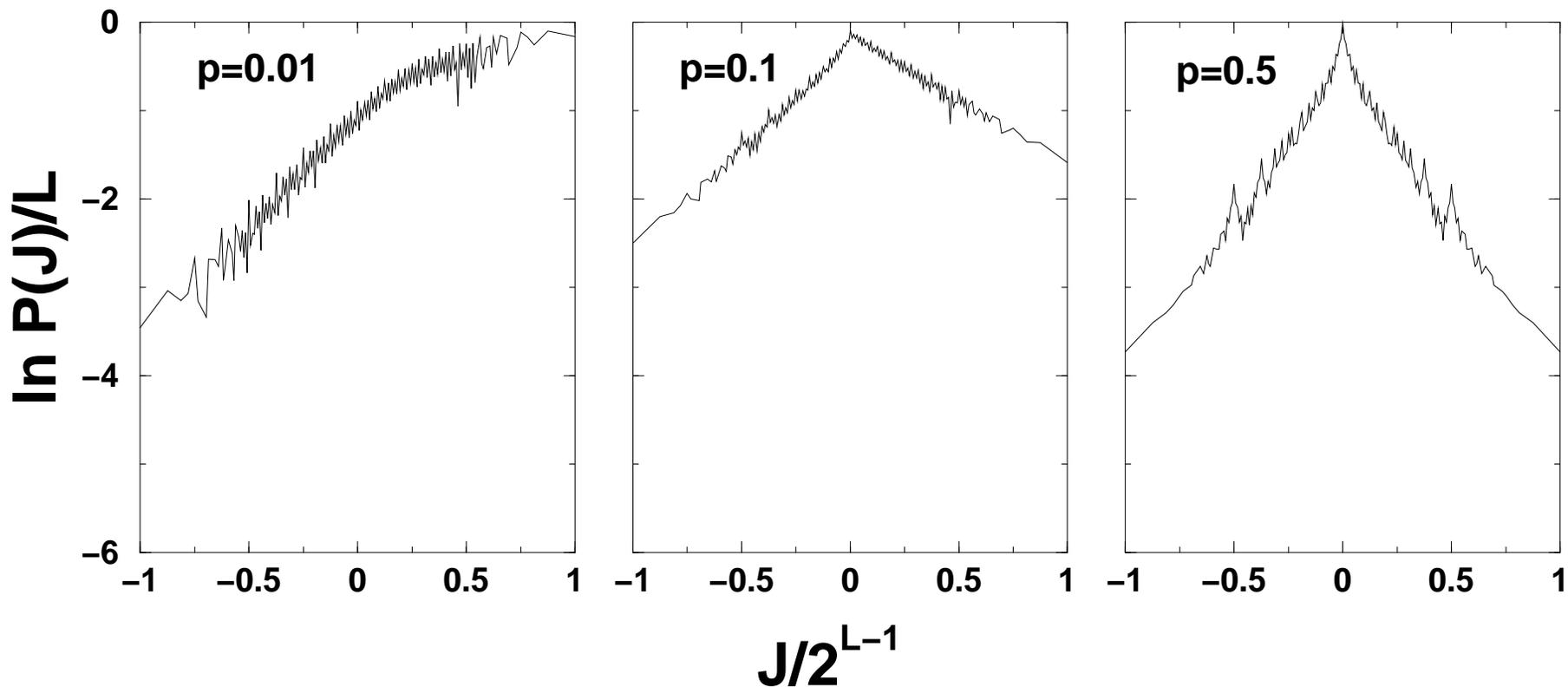
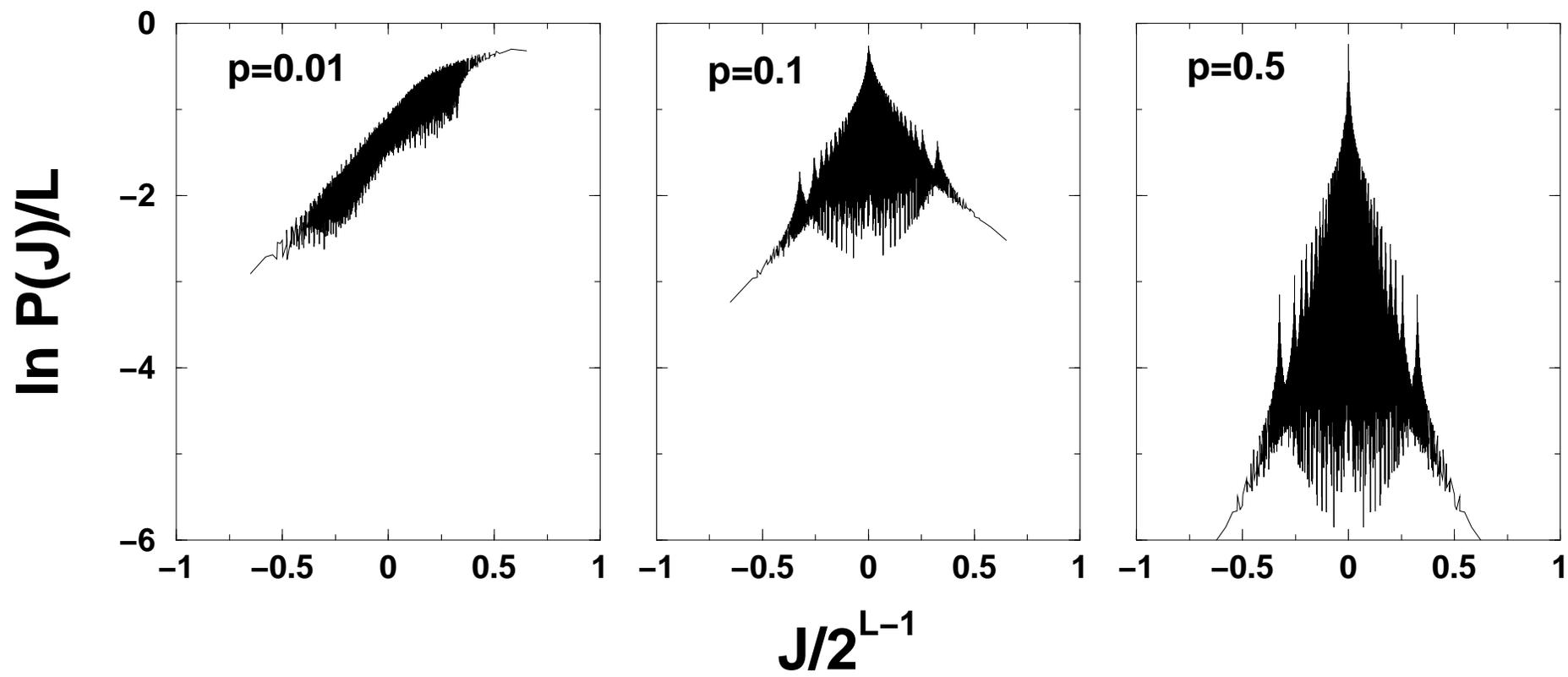
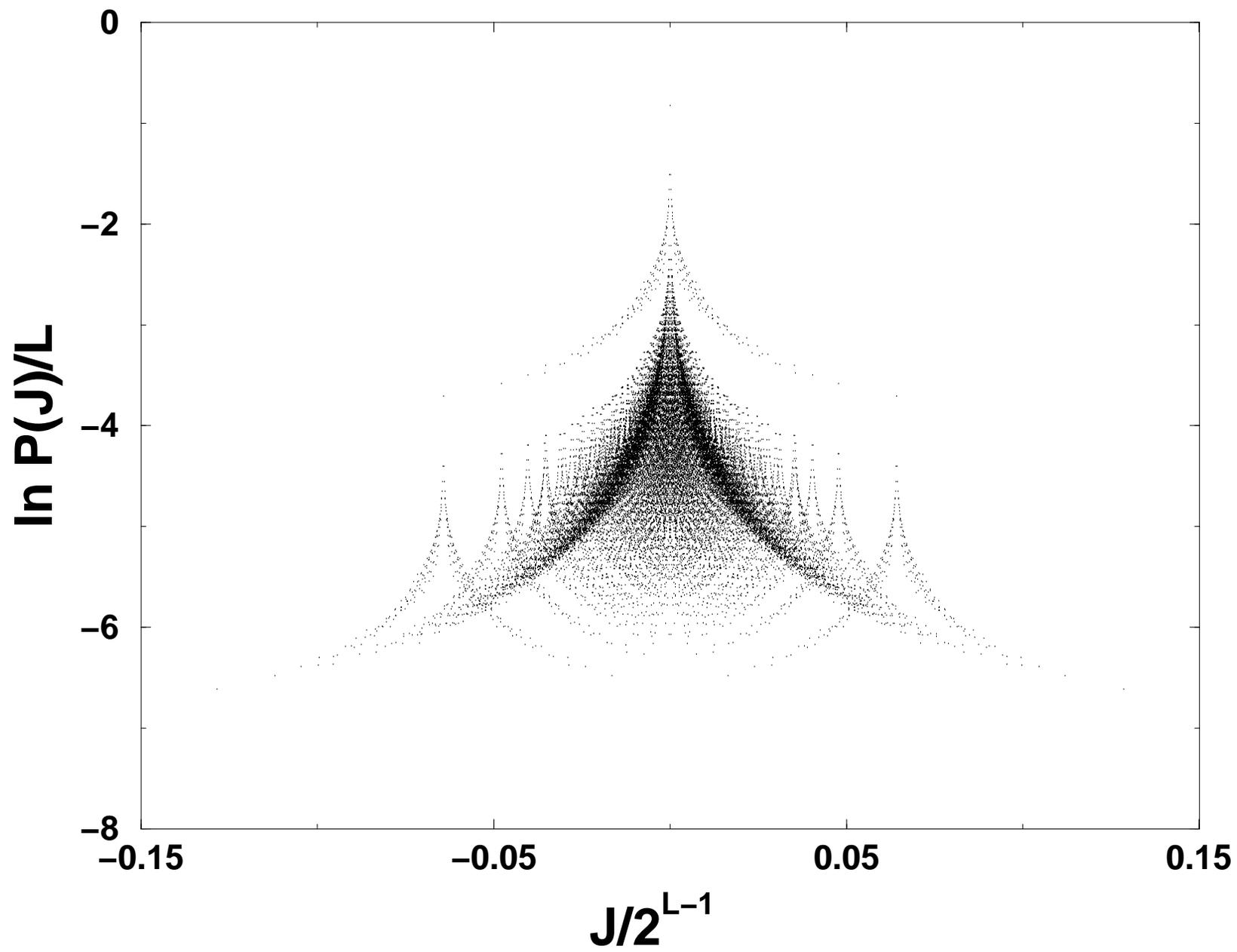
\end{document}